\documentstyle[aps,multicol,epsf]{revtex}

\catcode`\@=11

\def\eqbegin         {  \begin{eqnarray}  }
\def\eqend           {  \end{eqnarray}  }
\def\beq{\begin{equation}}
\def\eeq{\end{equation}}
\def\sectionnumbering
{\setcounter{equation}{0}\renewcommand{\theequation}
{\arabic{section}.\arabic{equation}}}

\def\del          { \partial }

\def\N{{\bf N}\ }
\def\Z{{\bf Z}}

\def\disp{\displaystyle}

\def\hs_2{\hspace{2mm}}
\def\hs_3{\hspace{3mm}}

\title{Persistent Edge Currents for Paired Quantum Hall States}
\author{Kazusumi Ino \thanks{e-mail:ino@nashi.issp.u-tokyo.ac.jp}}

\begin{document}
\maketitle
\begin{center}
{\it Nomura Research Institute} \\
{\it Hongo 2-2-9,Bunkyo-ku  ,  Tokyo,  113-0033, Japan}
\end{center}
\thispagestyle{empty}
\begin{abstract}
We study the behavior of the persistent edge current for paired quantum Hall states on the cylinder. We show that the currents are  periodic with the unit
flux $\phi_0=hc/e$.At low temperatures,
they exhibit anomalous  oscillations in
their flux dependence.
The shape of the functions converges to the sawtooth
function periodic with $\phi_0/2$.  \\
PACS: 73.40Hm, 74.20-z.
\end{abstract}

%\begin{multicols}{2}

\section{Introduction}
\sectionnumbering
One of the surprising aspects of the fractional quantum Hall
effect is that the edge state forms  a new kind of  state  of matter
beyond Fermi liquid, called the  chiral Tomonaga-Luttinger liquid \cite{wen}.
 Some experiments have already  demonstrated  the characteristic behavior of
chiral Tomonaga-Luttinger liquid \cite{kane,milli}.
Recently the Aharanov-Bohm effect (AB effect) in such  systems  were
studied  by and Geller et al
\cite{geller,geller2} and Chamon et al \cite{chamon}.
Especially, the latter authors predict new edge-current
 oscillations in the persistent edge current of
 the $\nu=\frac{1}{q}$ Laughlin state,
 which has
no amplitude reduction from disorder and thus results in
a universal non-Fermi-liquid temperature dependence.
Also the persistent current for the annulus Laughlin state was
recently investigated by Kettemann \cite{ketteman}. These
studies show the current is periodic with a unit flux quantum
$\phi_0=hc/e$ in agreement with the theorem of Byers and Young
\cite{byer}.  The period would be $q\phi_0$ if the quasiparticle
had a unit charge $e$ instead of a fractional charge $e/q$
\cite{gefen}.

Motivated by these recent studies,
 we investigated  the persistent currents
in paired fractional Hall states in Ref.\cite{ino}.
We showed that the currents are periodic with
the unit flux $\phi_0=hc/e$.
At low temperatures,  they exhibit
anomalous  oscillations in their flux dependence.
The shape of the functions
converges to the sawtooth function periodic with $\phi_0/2$.
In this paper, we extend these results to
 the states on the cylinder geometry and discuss some extremal limits.

The states we will consider are the 331 state \cite{halp},
the Haldane-Rezayi state \cite{haldane2}
 and the Pfaffian state \cite{moore}.  They are
quantum Hall analogs of the BCS superconductor.
The pairing
symmetry is p-wave with $S_z=1,0$ for  the Pfaffian and the 331 states
respectively, and
d-wave for  the Haldane-Rezayi states.
 The Haldane-Rezayi state and the Pfaffian state are
 candidate states for  $\nu =5/2$ plateau \cite{willet}.
 They are supposed to exhibit
 some novel features beyond ordinary quantum Hall states,
such as nonabelian statistics and
 specific degeneracy on a surface with nontrivial topology.
 On the other hand, the 331 state is a part of generalized hierarchy
\cite{wenzee}, but
 can be interpreted as a paired state.
  It is believed to be realized  at
$\nu=\frac{1}{2}$ plateau observed in double layer systems \cite{he}.

The organization of this paper is as follows. In Sec.2, we
recall the edge states of the paired quantum Hall states
In Sec.3, we deduce the exact formulas  of the persistent edge
currents and discuss their analytic properties.
We compare the persistent currents by numerical plots.
In Sec.4,
We consider some extremal limits and their effect on the persistent edge currents.
Sec.6 summarizes our conclusions. The definitions of
Jacobi $\theta$ functions are given in Appendix.

\section{Edge States for  Paired Quantum Hall States}
\sectionnumbering
Let us recall edge excitations on
the circumference $L$ of the $\nu=\frac{1}{q}$ paired quantum Hall states.

The charge sector  is
a chiral Luttinger liquid $\varphi$.
Its Hilbert space is generated by
 the $U(1)$ Kac-Moody algebra formed by
$j=\frac{1}{\sqrt{q}}\del \varphi$
 and the zero modes corresponding to quasiparticles \cite{wen}.
The Hamiltonian is given by
$H=\frac{1}{2}\sum_{n \in Z} j_{n}j_{-n}-\frac{c}{24}$
where we include a Casimir factor with $c=1$.
The complete description of the edge
excitations can be given  by  the grand partition function.
For the sector with the charge $r/q , \hspace{2mm} r =0,1\cdots,q-1$,
they are given by
\eqbegin
\chi_{r/q}^{\rm even}(\tau,\phi)&=&
\frac{1}{\eta}\sum_{m \in Z_{\rm even}}
e^{2\pi i\tau\frac{(mq+r)^2}{2q}+2\pi i\phi(m+\frac{r}{q})}
\label{r/q1}
\\
\chi_{r/q}^{\rm odd}(\tau,\phi)&=&
\frac{1}{\eta}\sum_{m \in Z_{\rm odd}}
e^{2\pi i\tau\frac{(mq+r)^2}{2q}+2\pi i\phi(m+\frac{r}{q})}
\label{r/q2}
\\
\\
\chi_{r/q}&=&\chi_{r/q}^{\rm even}+\chi^{\rm odd}_{r/q}.
\eqend
where even (odd) refers to the number of electrons and
$\eta$ is the Dedekind function:
\eqbegin
\eta(\tau)&=&x^{\frac{1}{24}}\prod_{n=1}^{\infty}(1-x^n), \label{dedekind}\\
x&=&e^{\disp 2\pi i \tau}.
\eqend
The finite size $L$ induces a temperature scale, which we will take as
\eqbegin
T_0=\frac{\hbar v}{ k_{B}L},
\eqend
where $v$ is the  Fermi velocity of the edge modes
determined by the confining potential.
For example, a Fermi velocity of $10^{6}$ cm/s and circumference of
$1\mu $ m yields $T_0 \sim  60$ mK. $\tau$ is given by $iT_0/T$.

Let us suppose that the edge state is coupled to an Aharanov-Bohm
flux $\Phi$.
The coupling of  the edge state to an Aharanov-Bohm flux $\Phi$
is achieved by putting  $\phi=\Phi/\phi_0$ to $\phi$ in
(\ref{r/q1})(\ref{r/q2})
with    $\phi_0=hc/e$,  the unit flux quantum.

Next we consider  the extra internal degrees of freedom other than
chiral Luttinger liquid.
In the bulk conformal field theory description,
pairing is due to  the internal degrees of freedom given by
some kind of  fermion $\psi$.  The charge degrees of freedom is given
by a chiral boson $\varphi$.  The operator for the electron
is  of the  form $\psi e^{i\sqrt{q}\varphi}$ where $q$ is even.
Thus the filling fraction $\nu=1/q$ have an even denominator.

The edge states of these states have a fermionic sector corresponding
to $\psi$.
The fermionic and charge  sectors are not
decoupled in the following sense:
the fermionic internal degree of freedom requires the global
selection rules in the bulk and edge of the paired state.
Thus the Hilbert spaces of the edge excitations of these states
 are not simply the direct sum of the sectors and
the grand partition functions are not given by  the direct product of
the ones for these sectors.

The additional sectors of edge excitations for
Pfaffian, Haldane-Rezayi, 331 states
 are  given by Majorana-Weyl fermions (MW),
symplectic fermions(Sf), Dirac fermions (D)
respectively.
These modes contribute to  the partition function
 through the following functions for
untwisted and twisted sectors with even or odd number of fermions :
\eqbegin
\chi^{\rm MW}_1(\tau)&=&\frac{1}{2}x^{-\frac{1}{48}}
\left(\prod_{0}^{\infty}(1+x^{n+\frac{1}{2}}) +
 \prod_{0}^{\infty}(1-x^{n+\frac{1}{2}})
 \right) \\
 &=&\frac{1}{2}\left(\sqrt{\frac{\theta_3(\tau)}{\eta(\tau)}}+
\sqrt{\frac{\theta_4(\tau)}{\eta(\tau)}}\right), \\
 \chi^{\rm MW}_{\psi}(\tau)&=&
 \frac{1}{2}x^{-\frac{1}{48}}\left(\prod_{0}^{\infty}(1+x^{n+\frac{1}{2}}) -
 \prod_{0}^{\infty}(1-x^{n+\frac{1}{2}})
 \right)\\
 &=&
 \frac{1}{2}\left(\sqrt{\frac{\theta_3(\tau)}{\eta(\tau)}}
 -\sqrt{\frac{\theta_4(\tau)}{\eta(\tau)}}\right), \\
 \chi^{\rm MW}_{\sigma}(\tau)&=&
 x^{\frac{1}{24}}\prod_{1}^{\infty}(1+x^n) \\
&=&\sqrt{\frac{\theta_2(\tau)}{2\eta(\tau)}}.
\eqend

\eqbegin
\chi_{1}^{\rm Sf}(\tau)
&=&
 \frac{1}{2}x^{\frac{1}{12}}\left(\prod_{1}^{\infty}(1+x^n)^2 +
\prod_{1}^{\infty}(1-x^n)^2 \right) \\
&=&\frac{1}{2}
\left(\frac{\theta_2(\tau)}{2\eta(\tau)}+\eta(\tau)^2\right),
\\
\chi_{\psi}^{\rm Sf}(\tau)&=&
\frac{1}{2}x^{\frac{1}{12}}\left(\prod_{1}^{\infty}(1+x^n)^2
- \prod_{1}^{\infty}(1-x^n)^2 \right) \\
&=&\frac{1}{2}
\left(\frac{\theta_2(\tau)}{2\eta(\tau)}-\eta(\tau)^2\right),
\\
\chi^{\rm Sf}_{\sigma}
&=&\frac{1}{2}x^{-\frac{1}{24}}
\left(\prod_{0}^{\infty}(1+x^{n+\frac{1}{2}})^2 +
 \prod_{0}^{\infty}(1-x^{n+\frac{1}{2}})^2
 \right) \\
 &=&\frac{1}{2}\left(\frac{\theta_3(\tau)}{\eta(\tau)}
+\frac{\theta_4(\tau)}{\eta(\tau)}\right),
\\
\chi^{\rm Sf}_{\widetilde{\sigma}}
&=&
 \frac{1}{2}x^{-\frac{1}{24}}\left(\prod_{0}^{\infty}(1+x^{n+\frac{1}{2}})^2 -
 \prod_{0}^{\infty}(1-x^{n+\frac{1}{2}})^2
 \right)\\
 &=&
 \frac{1}{2}\left(\frac{\theta_3(\tau)}{\eta(\tau)}
 -\frac{\theta_4(\tau)}{\eta(\tau)}\right).
\eqend

\eqbegin
\chi^{\rm D}_1 &=&\frac{1}{2}x^{-\frac{1}{24}}
\left(\prod_{0}^{\infty}(1+x^{n+\frac{1}{2}})^2 +
 \prod_{0}^{\infty}(1-x^{n+\frac{1}{2}})^2
 \right) \\
 &=&\frac{1}{2}\left(\frac{\theta_3(\tau)}{\eta(\tau)}
+\frac{\theta_4(\tau)}{\eta(\tau)}\right), \\
\chi^{\rm D}_{\psi}&=&
 \frac{1}{2}x^{-\frac{1}{24}}
\left(\prod_{0}^{\infty}(1+x^{n+\frac{1}{2}})^2 -
 \prod_{0}^{\infty}(1-x^{n+\frac{1}{2}})^2
 \right)\\
 &=&
 \frac{1}{2}\left(\frac{\theta_3(\tau)}{\eta(\tau)}
 -\frac{\theta_4(\tau)}{\eta(\tau)}\right),\\
 \chi^{\rm D}_{\sigma}
&=&
 \frac{1}{2}x^{\frac{1}{12}}\left(\prod_{1}^{\infty}(1+x^n)^2+\prod_{1}^{\infty}(1-x^n)^2 \right)\\
&=&\frac{1}{2}\left(\frac{\theta_2(\tau)}{2\eta(\tau)}-\eta(\tau)^2\right),
\\
 \chi^{\rm D}_{\widetilde{\sigma}}&=&
 \frac{1}{2}x^{\frac{1}{12}}\left(
 \prod_{1}^{\infty}(1+x^n)^2-\prod_{1}^{\infty}(1-x^n)^2 \right)\\
&=&\frac{1}{2}\left(\frac{\theta_2(\tau)}{2\eta(\tau)}-\eta(\tau)^2\right).
\eqend

\section{Persistent edge current for the states on the cylinder}
Let us derive the persistent edge current for the state on
the cylinder.
For the Laughlin state, the sum of the edge currents
(for the annulus) was studied by Ketteman some \cite{ketteman}.
In this section, we study the sum of the currents for
Pfaffian, 331 and Haldane-Rezayi state.

The AB flux is through the center of the annulus.
The low-energy excitations are edge excitations on the inner
and our edges. On the bulk, it is necessary to consider
the grand-canonical ensemble of the electrons, which leads
to the constraints between the excitation on the two edges.
The grand-canonical partition functions emerging here
are the ones for the corresponding extended
conformal field theories  with the electron as the extending field.
The grand-canonical partition functions with AB flux are
\cite{milo,ino2}
\\
{\it Pfaffian State}
\eqbegin
Z_{\rm Pf}^{\rm cyl}(\tau,\phi)=
=\sum_{r=0}^{q-1}\left[ |\chi_1^{\rm MW}\chi_{r/q}^{\rm even}+
\chi_{\psi}^{\rm MW}\chi_{r/q}^{\rm odd}|^2+
|\chi_{\psi}^{\rm MW}\chi_{r/q}^{\rm even}+
\chi_1^{\rm MW}\chi_{r/q}^{\rm odd}|^2+
|\chi_{\sigma}\chi_{(r+1/2)/q}|^2  \right].
\eqend
{\it 331 State}
\eqbegin
Z_{\rm 331}^{\rm cyl}(\tau,\phi) =
\sum_{r=0}^{q-1}\left[ |\chi_1^{\rm D}\chi_{r/q}^{\rm even}+
\chi_{\psi}^{\rm D}\chi_{r/q}^{\rm odd}|^2+
|\chi_{\psi}^{\rm D}\chi_{r/q}^{\rm even}+
\chi_1^{\rm D}\chi_{r/q}^{\rm odd}|^2+
2|\chi^{\rm D}_{\sigma}\chi_{(r+1/2)/q}|^2  \right]
\eqend
{\it Haldane-Rezayi State} \\
\eqbegin
Z_{\rm HR}^{\rm cyl}(\tau,\phi)=
\sum_{r=0}^{q-1}\left[ |\chi_1^{\rm Sf}\chi_{r/q}^{\rm even}+
\chi_{\psi}^{\rm Sf}\chi_{r/q}^{\rm odd}(\tau,\phi)|^2+
|\chi_{\psi}^{\rm Sf}\chi_{r/q}^{\rm even}+
\chi_1^{\rm Sf}\chi_{r/q}^{\rm odd}|^2+
|\chi^{\rm Sf}_{\sigma}\chi_{(r+1/2)/q}|^2  \right].
\label{anHR}
\eqend

By using the modular transformation, we end up with  the following
expressions in terms of theta functions ($t$ is  $-\frac{1}{\tau}$),
\eqbegin
Z_{\rm Pf}&=&\frac{1}{2q\eta(t)^4}
\sum_{r=0}^{q-1}\Bigl[ \left|\sqrt{\theta_3(0|t)}\theta_3(\frac{\phi+r}{q}|\frac{t}{q})\right|^2
+\left|\sqrt{\frac{\theta_4(0|t)}{2}}\theta_3(\frac{\phi+r+1/2}{q}|\frac{t}{q})\right|^2
+\left|\sqrt{\theta_2(0|t)}\theta_1(\frac{\phi+r}{q}|\frac{t}{q})\right|^2 \Bigr]. \\
Z_{\rm 331}&=&\frac{1}{2q\eta(t)^4}\sum_{r=0}^{q-1}\Bigl[
\left|\theta_3(0|t)\theta_3(\frac{\phi+r}{q}|\frac{t}{q})\right|^2
+\left|\theta_4(0|t)\theta_3(\frac{\phi+r+1/2}{q}|\frac{t}{q})\right|^2
+\left|\theta_2(0|t)\theta_1(\frac{\phi+r}{q}|\frac{t}{q})\right|^2 \Bigr]. \\
Z_{\rm HR}&=&\frac{1}{2q\eta(t)^4}\sum_{r=0}^{q-1}\Bigl[
\left|\theta_4(0|t)\theta_3(\frac{\phi+r}{q}|\frac{t}{q})\right|^2
+\left|\theta_3(0|t)\theta_3(\frac{\phi+r+1/2}{q}|\frac{t}{q})\right|^2
+\left|\theta_2(0|t)\theta_2(\frac{\phi+r}{q}|t/q)\right|^2 \Bigr]. \\
\eqend
The formulae for the sum of the persistent currents on
two edges are  obtained by differentiation  of
the free energy $-T{\rm log}Z$ in term of $\phi$
\eqbegin
I\equiv \frac{T}{\phi_0}
\frac{\del {\rm ln}Z(\tau,\phi)}{\del{\phi}}.
\label{perfor}
\eqend

\begin{figure}[htbp]
\vspace{.2cm}
\noindent
\hspace{1.5 in}
\epsfxsize=3.5in
\epsfbox{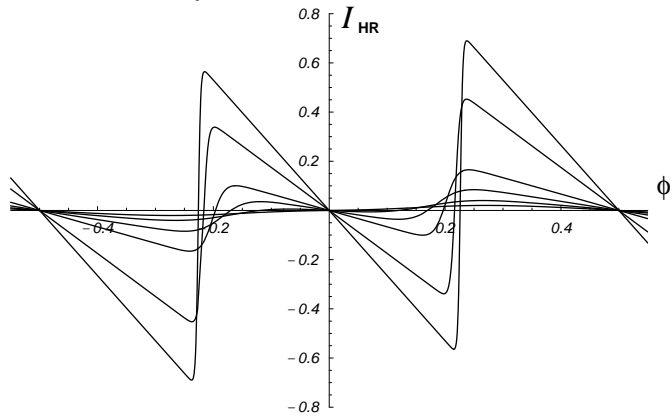}
\vspace{.5cm}
\caption[flux1_2]
{The flux dependence of the total persistent current
 for the $\nu=1/2$ Haldane-Rezayi state on the cylinder
  at temperatures $T/T_0=0.26,0.23,0.2,0.17,0.15,0.135$.}
\label{HRcylinder}
\end{figure}

We plot in {\bf Fig.\ref{HRcylinder}}
the flux dependence of the persistent
edge current for $\nu=1/2$ Haldane-Rezayi state
 at temperatures $T/T_0=0.26,0.23,0.2,0.17,0.15,0.135$.
It shows a similar low-temperature
behavior as in the case of disk.
This is interpreted as an incorporation
between the Byers-Yang and Mermin-Wagner theorems.
The Pfaffian and 331 states also show a similar behavior but
different.
At low temperature limit, they converge as
\eqbegin
I &\rightarrow&
-2\nu \frac{ev}{k_BL}(\phi-\frac{1}{2}r),
\hspace{4mm}  -\frac{1}{4}+\frac{1}{2}r<\phi< \frac{1}{4}+\frac{1}{2}r, \hspace{4mm} r \in \Z.
\eqend
Note that the total amount of the current is as twice as  the one in the disk case.

Anomalous oscillations of the persistent currents
can be  explained  from the pairing of composite fermions in the bulk and edge
state of paired states.
Naively, the edge persistent currents
which we have calculated should have a period $\phi_0/2$
 since the  bulk states of paired states are in a BCS
superconducting phase and the order parameter has a charge $2e$.
However, from the Mermin-Wigner theorem,
the edge states can not be a BCS-like condensate except
at zero temperature.
Then, from the behavior of the currents we have found, we see
that  as we lower the temperature, the edge states become
 closer to a BCS-like condensate,  but the Mermin-Wigner theorem
prevents the phase transition of the edge states.
The effect of the pairing of composite fermions in the bulk
changes continuously as a function of the temperature.
The BCS pairing  in the edge states  becomes
 stronger   at lower temperatures,  but the condensation
 only occurs at zero temperature.

This phenomenon may be seen as   an interesting  bridge
between superconductivity in 2+1 dimensions
 and superconductivity in 1+1 dimensions.

Thus we  see that the predicted behavior
 of  persistent edge currents can be used as
a  method to distinguish the bulk topological order.
Especially the flux dependence
can reveal the pairing of composite fermions in the bulk of
fractional quantum Hall states.

Experimentally, the magnitude of the persistent currents is
 preferably measurable at low temperatures and small samples.
As we have predicted, experiments at the even-denominator
plateaus may detect the anomalous oscillations
of the persistent current.

\section{From cylinder to disk}
By considering different radii $R_1$, $R_2$
for two edges respectively in the formulas
in the previous section, the persistent edge
current for the annulus case can be treated.
In this case,  different temperature scales
$T_1=\frac{\hbar v}{ 2\pi k_{B}R_1}$ and
$T_2=\frac{\hbar v}{ 2\pi k_{B}R_2}$
 arise.
We will consider some extremal limits in this section.
The extremal limits we will consider is the
low temperature limit and the small radius limit.
We will see that the disk grand partition function
play an important role.

Let us consider the low temperature behavior of
persistent current for a FQH on a cylinder.
Generally the grand-partition function $Z(\tau,\phi)$ for
a FQH on a cylinder has a form
\eqbegin
Z(\tau,\phi) = \sum_{i,j}  N_{ij}\overline{Z_{i}} Z_{j}
\eqend
where $i$ and $j$ are indices for each sector in
the radii $R_1$ and $R_2$ edge states respectively
and $N_{ij}$ are integer coefficients.

As we change $T$ to  $0$, for a given value of $\phi$,
the sector determined by $\phi$ gives a dominant
contribution to $Z$.  We put it as $i$-th sector, then
the partition function at $T \sim 0$ becomes
\eqbegin
Z(\tau,\phi) \sim \overline{Z_{i}}\sum_{j}N_{ij} Z_{j}.
\eqend
This decomposition is valid except around $\phi=\pm 1/4$.
Thus the persistent edge current is divided into a sum
\eqbegin
I \sim \overline{I_1}  + I_2,
\eqend
where $\overline{I_1},I_2$ is calculated only from the one edge and
actually same as the one calculated from
the disk partition function $Z_{\rm disk} = \sum_{i,j}N_{ij}Z_j$.
Thus at sufficiently low temperature, the total
persistent edge current is calculated from the one for
the disk case except around $\phi=\pm 1/4$.

It is also interesting to
 consider an extremal limit $R_1 \rightarrow 0$
 with the AB flux fixed at $\phi$.
As we change $R_1$ to  $0$,
$T_1$ goes to large. Thus for the
small radius edge, we are effectively in the low
temperature region.
Thus the sector determined by $\phi$ gives a dominant contribution to $Z$.
As above, the partition function at $R_1 \sim 0$ becomes
\eqbegin
Z(\tau,\phi) \sim \overline{Z_{i}}\sum_{j}N_{ij} Z_{j},
\eqend
except around $\phi=\pm 1/4$.
Thus the persistent edge current is divided into a sum
\eqbegin
I \sim \overline{I_1}  + I_2,
\eqend
where $I_2$ can be calculated from the disk partition function
$Z_{\rm disk}$ except around $\phi=\pm 1/4$.

Thus at small radius,
the persistent edge current on the one edge
can be calculated from the formula for the disk case.
This observation can be extended to a FQH state on a region with
arbitrary number of boundaries.

\section{Conclusions}
In this paper, we have deduced the following conclusions extending
our previous result for the disk case:
\begin{enumerate}
\item
The total amount of
persistent edge  currents in the paired quantum Hall states on the cylinder
are flux periodic with
period $\phi_0$ in agreement with the theorem of Byer and Young.

\item
At low temperatures, the currents exhibit
anomalous oscillations as  functions of
the Aharanov-Bohm flux. The shape of
anomalous oscillations depends on
the bulk topological order and converges to
a sawtooth function with
 period  $\phi_0/2$ at zero temperature.

\item
This behaviors  implies
that, although the bulk states of paired states are in a
BCS superconducting phase,  the edge states are
a BCS condensate only at zero temperature.
The BCS pairing structure changes the flux dependence
of  the persistent current continuously and there is no
phase transition at finite temperatures.
This is in agreement with the Mermin-Wigner theorem.

\item
These pairing effects, especially anomalous oscillations,
 may provide a  means to observe paired quantum Hall states at
the $\nu=5/2$ plateau and the $\nu=1/2$ plateau in double layers.

\item
The phenomenon we predict may be seen as  an interesting
bridge between the superconductivity in 2+1 dimensions and
in  1+1 dimensions.

\end{enumerate}

{\it Acknowledgement.} The author would like to thank
T.Ando, D.Lidsky, M.Kohmoto, J.Shiraishi
for useful discussions and M.R.Geller  and especially M.Flohr
for correspondence.

\vskip 0.6in

\appendix
{\Large \bf Appendix: Jacobi $\theta$ functions}
\vskip 0.3in
\sectionnumbering

The Jacobi $\theta$ functions $\theta_1, \theta_2, \theta_3,\theta_4$
are examples of
modular forms.  They are defined by
\eqbegin
\theta_1(\zeta |  \tau) &=&
-i\sum_{r \in \Z+1/2}(-1)^{r-1/2}y^r q^{r^2/2}, \\
\theta_2(\zeta |  \tau) &=& \sum_{r \in \Z+1/2}y^r q^{r^2/2}, \\
\theta_3(\zeta |  \tau) &=& \sum_{n \in \Z}y^n q^{n^2/2}, \\
\theta_4(\zeta |  \tau) &=& \sum_{n \in \Z}(-1)^n y^n q^{n^2/2},
\eqend
with $q={\rm exp}2\pi i \tau$ and $y={\rm exp}2\pi i\zeta$.
They also have the following infinite product forms:
\eqbegin
\theta_1(\zeta |  \tau) &=&
-iy^{1/2}q^{1/8}\prod_{n=1}^{\infty}(1-q^n)
\prod_{n=0}^{\infty}(1-yq^{n+1})(1-y^{-1}q^n), \\
\theta_2(\zeta|  \tau) &=&
y^{1/2}q^{1/8}\prod_{n=1}^{\infty}(1-q^n)
\prod_{n=0}^{\infty}(1+yq^{n+1})(1+y^{-1}q^n), \\
\theta_3(\zeta |  \tau) &=& \prod_{n=1}^{\infty}(1-q^n)
\prod_{r\in \N+1/2}(1+yq^{r})(1+y^{-1}q^r), \\
\theta_4(\zeta |  \tau) &=& \prod_{n=1}^{\infty}(1-q^n)
\prod_{r\in \N+1/2}(1-yq^{r})(1-y^{-1}q^r).
\eqend
The differential of $\theta(\zeta|\tau)$
with  respect to $\zeta$ are denoted as  $\theta'(\zeta | \tau)$.

The transformation property of Jacobi theta functions and
the Dedekind $\eta$ function Eq.(\ref{dedekind} ) under $S$ is
\eqbegin
\theta_2(-1/\tau)=\sqrt{-i\tau}\theta_4(\tau), \\
\theta_3(-1/\tau)=\sqrt{-i\tau}\theta_3(\tau), \\
\theta_4(-1/\tau)=\sqrt{-i\tau}\theta_2(\tau), \\
\eta(-1/\tau)  = \sqrt{-i\tau} \eta(\tau)\ .
\eqend

\def\NP{{\it Nucl. Phys.\ }}
\def\PRL{{\it Phys. Rev. Lett.\ }}
\def\PL{{\it Phys. Lett.\ }}
\def\PR{{\it Phys. Rev.\ }}
\def\CMP{{\it Comm. Math. Phys.\ }}
\def\IJMP{{\it Int. J. Mod. Phys.\ }}
\def\MPL{{\it Mod. Phys. Lett.\ }}
\def\RMP{{\it Rev. Mod. Phys.\ }}
\def\AP{{\it Ann. Phys. (NY)\ }}
%

%\end{multicols}

\newpage

\end{document}